\def\presentation{
\voffset -.50in  \hoffset -.19in
\oddsidemargin 0in \evensidemargin 0in
\marginparwidth .75in \marginparsep 7pt \topmargin 0in
\headheight 12pt \headsep .25in
\footheight 18pt \footskip .35in
\textheight 9.5in \textwidth 6.5in
\columnsep 10pt \columnseprule 0pt }
\documentstyle{article}
\presentation
\begin{document}
%

%
\def\tilde{\widetilde}
\def\bar{\overline}
\def\hat{\widehat}
\def\*{\star}
\def\[{\left[}
\def\]{\right]}
\def\({\left(}
\def\){\right)}
\def\zb{{\bar{z} }}
\def\frac#1#2{{#1 \over #2}}
\def\inv#1{{1 \over #1}}
\def\half{{1 \over 2}}
\def\d{\partial}
\def\der#1{{\partial \over \partial #1}}
\def\dd#1#2{{\partial #1 \over \partial #2}}
\def\vev#1{\langle #1 \rangle}
\def\bra#1{{\langle #1 |  }}
\def\ket#1{ | #1 \rangle}
\def\rvac{\hbox{$\vert 0\rangle$}}
\def\lvac{\hbox{$\langle 0 \vert $}}
\def\2pi{\hbox{$2\pi i$}}
\def\e#1{{\rm e}^{^{\textstyle #1}}}
\def\grad#1{\,\nabla\!_{{#1}}\,}
\def\dsl{\raise.15ex\hbox{/}\kern-.57em\partial}
\def\Dsl{\,\raise.15ex\hbox{/}\mkern-.13.5mu D}
\def\comm#1#2{ \BBL\ #1\ ,\ #2 \BBR }
\def\x{\stackrel{\otimes}{,}}
\def\det{ {\rm det}}
\def\tr{{\rm tr}}
%
%
\def\th{\theta}		\def\Th{\Theta}
\def\ga{\gamma}		\def\Ga{\Gamma}
\def\be{\beta}
\def\al{\alpha}
\def\ep{\epsilon}
\def\la{\lambda}	\def\La{\Lambda}
\def\de{\delta}		\def\De{\Delta}
\def\om{\omega}		\def\Om{\Omega}
\def\sig{\sigma}	\def\Sig{\Sigma}
\def\vphi{\varphi}
%
%
\def\CA{{\cal A}}	\def\CB{{\cal B}}	\def\CC{{\cal C}}
\def\CD{{\cal D}}	\def\CE{{\cal E}}	\def\CF{{\cal F}}
\def\CG{{\cal G}}	\def\CH{{\cal H}}	\def\CI{{\cal J}}
\def\CJ{{\cal J}}	\def\CK{{\cal K}}	\def\CL{{\cal L}}
\def\CM{{\cal M}}	\def\CN{{\cal N}}	\def\CO{{\cal O}}
\def\CP{{\cal P}}	\def\CQ{{\cal Q}}	\def\CR{{\cal R}}
\def\CS{{\cal S}}	\def\CT{{\cal T}}	\def\CU{{\cal U}}
\def\CV{{\cal V}}	\def\CW{{\cal W}}	\def\CX{{\cal X}}
\def\CY{{\cal Y}}	\def\CZ{{\cal Z}}
%
%
\font\numbers=cmss12
\font\upright=cmu10 scaled\magstep1
\def\stroke{\vrule height8pt width0.4pt depth-0.1pt}
\def\topfleck{\vrule height8pt width0.5pt depth-5.9pt}
\def\botfleck{\vrule height2pt width0.5pt depth0.1pt}
\def\Zmath{\vcenter{\hbox{\numbers\rlap{\rlap{Z}\kern
		0.8pt\topfleck}\kern
		2.2pt \rlap Z\kern 6pt\botfleck\kern 1pt}}}
\def\Qmath{\vcenter{\hbox{\upright\rlap{\rlap{Q}\kern
                   3.8pt\stroke}\phantom{Q}}}}
\def\Nmath{\vcenter{\hbox{\upright\rlap{I}\kern 1.7pt N}}}
\def\Cmath{\vcenter{\hbox{\upright\rlap{\rlap{C}\kern
                   3.8pt\stroke}\phantom{C}}}}
\def\Rmath{\vcenter{\hbox{\upright\rlap{I}\kern 1.7pt R}}}
\def\Z{\ifmmode\Zmath\else$\Zmath$\fi}
\def\Q{\ifmmode\Qmath\else$\Qmath$\fi}
\def\N{\ifmmode\Nmath\else$\Nmath$\fi}
\def\C{\ifmmode\Cmath\else$\Cmath$\fi}
\def\R{\ifmmode\Rmath\else$\Rmath$\fi}
\def\cadremath#1{\vbox{\hrule\hbox{\vrule\kern8pt\vbox{\kern8pt
			\hbox{$\displaystyle #1$}\kern8pt} 
			\kern8pt\vrule}\hrule}}
\def\proof{\noindent {\underline {Proof}.} }
\def\cqfd{ {\hfill{$\Box$}} }
\def\square{ {\hfill \vrule height6pt width6pt depth1pt} \vspace{8pt}} 
%
%
\def\debut{ \begin{eqnarray} }
\def\fin{ \end{eqnarray} }
\def\non{ \nonumber }
%

%
%
\rightline{SPhT-99-017}
\vskip 1cm
\centerline{\LARGE Vertex Operator Solutions of}
\bigskip
\centerline{\LARGE  2d Dimensionally Reduced Gravity.}
\vskip 1cm

\vskip1cm

\centerline{\large  Denis Bernard
\footnote[1]{Member of the CNRS; dbernard@spht.saclay.cea.fr}
	and Nicolas Regnault 
\footnote[2]{regnault@spht.saclay.cea.fr} }
\centerline{Service de Physique Th\'eorique de Saclay
}
\centerline{F-91191, Gif-sur-Yvette, France.}

\vskip2cm

Abstract.\\
We apply algebraic and vertex operator techniques to solve two dimensional
reduced vacuum Einstein's equations. This leads to explicit expressions for 
the coefficients of metrics solutions of the vacuum equations as ratios of 
determinants. No quadratures are left. These formulas rely on the 
identification of dual pairs of vertex operators corresponding to
dual metrics related by the Kramer-Neugebauer symmetry.

\vfill
\newpage
%
%
\def\vv#1{ \vev{ #1 }_{z,\rho} }

\section{Introduction.}
Surprisingly, four dimensional gravity admits an integrable sector.
It corresponds to ansatz metrics which admit two surface orthogonal
Killing vectors. They describe stationary axis symmetric situations or 
colliding gravitational waves depending on the nature of the
Killing vectors. For these ansatz, the vacuum Einstein's equations reduce
to the Ernst's equations \cite{ernst}. An infinite dimensional solution
generating group for them was constructed by Geroch \cite{Geroch}
and later identified by Julia \cite{julia} as the affine $SL(2,R)$
Kac-Moody group. The integrable character of Ernst's equations was
deciphered by Belinskii-Zakharov in ref.\cite{Zakha}. See also ref.\cite{mais} for
a discussion relating these two aspects. Since then, various methods have been
applied to solve 2d reduced Einstein's equations: either using
Backlund or solution generating transformations or the Belinskii-Zakharov method,
cf eg. \cite{KhanPen,Nuktu,ChanXan,Ergarcia,letel,rerefs} for a sample of references
and \cite{Grif,Kramer} and references therein,
or using analytical finite-gap techniques \cite{gap}.
However, the by-now standard vertex operator approach to integrable
models developed by the Kyoto school \cite{kyoto} was never applied 
to this problem. This is probably due to the fact that the 
Belinskii-Zakharov method involves the so-called moving poles 
which forbids a direct application of the vertex operator technique. 
This problem was partially overcome in ref.\cite{BeJu}. The aim of this
paper is to fill the gap left open in this work and to describe how
vertex operators may be used to solve Einstein's equations.
It leads to determinant formula for the metrics which are described in the
following section. This vertex operator approach, which is based on an 
algebraic formulation of the dressing group, may also be useful 
for deciphering the Lie-Poisson properties of the
solution generating groups of Einstein's equations.
See refs.\cite{sem,BBis} for a discussion of the Lie-Poisson properties 
of the dressing transformations and ref.\cite{Sam} for a discussion
of these properties for the Geroch group. 
In supergravity context these groups are called duality groups 
and they are important for quantization. 

Besides providing explicit formulas for exact solutions of Ernst's
equations, (but whose possible physical applications are not discussed),
one of the aim of this paper is to decipher the algebraic
structures underlying the solvability of Ernst's equation.
Contrary to the impression that one may get from the
Belinskii-Zakharov approach which uses space-time dependent
spectral parameters, we shall show that Ernst's equations belong
to the usual class of integrable hierarchies, such as the
KP or sine-Gordon equations, and that they may be solved using usual
algebraic technical tools. Vertex operators is one of those techniques
which provide tools for algebraically solving matrix
Riemann-Hilbert problems. Deciphering these algebraic structures
leads us to identify dual pairs of elements of an affine Kac-Moody
group corresponding to solutions paired via the Kramers-Neugebauer
duality. This could be interesting in view of the importance
of the duality group in supergravity analogues of Ernst's equations.
But the long term motivation for this algebraic detours is quantization.
2D reduced gravity provides a toy model in application
to quantization technique to gravity. The dressing method, on
which our approach is based, is particularly adapted to quantization
since the dressing group usually acts on the phase space of integrable
hierachies by a Lie-Poison action. So it is promotted to a 
quantum group symmetry after quantization.
The next step in that direction would thus be to find a concise description 
of the symplectic struture of this system and of the Lie-Poisson property 
of the dressing group.
\bigskip

We have tried to write the paper such that it may be read in two different
ways depending if one is only interested in explicit formulas
for the solutions or if one is willing to learn the algebraic
structures underlying the derivation of these formulas.
Readers who are just interested in concrete formulas to
obtain solutions may restrict their attention to Section 2, 
beginning of Section 3 and Section 6. Algebraically oriented readers
may look at Section 4 and Section 5. The main new trick which
allows us to complete the approach initiated in \cite{BeJu}
is a construction of dual pairs of elements in the $SL(2,R)$ 
Kac-Moody group associated to dual pairs of solutions exchangeable
by Kramers-Neugebauer involution.
This leads us to solve a problem of factorization that was left opened 
in \cite{BeJu} and to give determinant formulas for the metrics. 

The vertex operator and the determinant formula of the metrics 
are described in the following Section 2.
The rest of the paper is devoted to the proof of these formulas
and it is  organized as follows.
In Section 3, we recall basic facts concerning the 2d reduced
Einstein's equations and present them in a way convenient for 
the following. In particular we introduce the appropriate
tau functions. Section 4 is devoted to the construction of
dual pairs of vertex operators which allow us to compute the dual 
pairs of tau functions. Section 5 describe how to get the coefficients
of the metric given the dual pair of vertex operators.
Finally, a few explicit examples and comparisons with
the previous result are described in Section 6.
Appendix A presents a rapid survey of the method described in \cite{BeJu}, 
and we have gathered a few useful formulas in Appendices B and C.

\section{Determinant formula for the metrics.}

As is well known, solutions of the  reduced Einstein's equations 
come in pairs which are related via the Kramer-Neugebauer duality.
The two dual metrics, that we shall denote by $ds^2$ and $ds_*^2$,
can be parametrized in terms of Weyl coordinates $z$ and $\rho$ as:
\debut
ds^2 &=&  \rho^{-\half}\, e^{2\hat \sig}~ (dz^2-d\rho^2)
+ G_{ab} dx^adx^b  \label{ds2}\\
ds^2_* &=&  \rho^{-\half}\, e^{2\hat \sig^*} ~ (dz^2-d\rho^2)
+ G^*_{ab} dx^adx^b \label{ds2dual}
\fin
All fields only depend on the two coordinates $z,\rho$.
The indices $a,b$ run from one to two and 
$\rho^2=\det G_{ab} = \det G^*_{ab}$.
The prefactor $\hat \sig$, or $\hat \sig^*$, is usually
called the conformal factor. The precise form
of the duality relation mapping $ds^2$ into $ds^2_*$
is recalled in the following section 
as well as an alternative parametrization of the metrics.

The components of the metrics would be parametrized in terms
of expectation values of certain vertex operators as follows:
\debut
e^{2\hat \sig} = |\vv{g}|^2 \quad &;&\quad
e^{2\hat \sig^*} = |\vv{g^*}|^2 \label{sig0}\\
e^{2\hat \sig} G_{22} = \sqrt{\rho}\ |\vv{g^*}|^2
\quad &;&\quad
e^{2\hat \sig^*} G^*_{22} = \sqrt{\rho}\ |\vv{g}|^2 \label{g22}\\
e^{2\hat \sig} G_{12} = \sqrt{\rho}~{\rm Im}\[{ \vv{g^*}\, \bar {\vv{Wg^*}} }\] 
\quad &;&\quad
e^{2\hat \sig^*} G^*_{12} = \sqrt{\rho}\ {\rm Im}\[{ \vv{g}\, \bar {\vv{Wg}} }\]
 \label{g12}\\
e^{2\hat \sig} G_{11} = \sqrt{\rho}\  |\vv{Wg^*}|^2
\quad &;&\quad
e^{2\hat \sig^*} G^*_{11} = \sqrt{\rho}\ |\vv{Wg}|^2 \label{g11}
\fin
Here $g$, $Wg$ and $g^*$, $Wg^*$ denote the vertex operators associated
to the two dual solutions. 
See eqs.(\ref{vertex},\ref{vertexdual}) and 
eqs.(\ref{wg},\ref{wgstar}) below for their definitions.
The overbar denotes complex conjugation.
The indices $z,\rho$ are here to recall that these expectation 
values depend on the Weyl coordinates.
Notice the interplay between $g$ and $g^*$ in the above formula: 
the element $g$ enters in the conformal factor $\hat \sig$ while 
the dual element $g^*$ enters in the metric $G_{ab}$.

The solutions we shall describe depend on two sets of
real parameters with total number $2(m+n)$: 
the first set is made of $2m$ parameters $(z_p, u_p)$, $p=1,\cdots,m$,
while the second set is made of $2n$ parameters denoted $(z_j,y_j)$,
$ j=1,\cdots,n$.
All parameters are real but $|z_j|>1$ and $|z_p|>1$.
It is convenient to introduce the notations $X_j,\mu_j$ and 
$X_p,\mu_p$ such that:
\debut
X_j =  \rho\ \frac{z_j^2-1}{(z_j-z)^2+\rho^2}
\quad {\rm and} \quad
\mu_j^2 &=& \frac{(z_j-z)+\rho}{(z_j-z)-\rho} \label{defY}
\fin
and similarly for $X_p$ and $\mu_p$. The functions $\mu_j$ are
related in a simple way to the usual moving poles
in the Belinskii-Zakharov approach. 
The validity of the metric is restricted to 
$(z,\rho)$ domains such that
$z\pm \rho < z_j$ for $z_j>1$ and
$z\pm \rho > z_j$ for $z_j<-1$, and
similarly for $z_p$.

The explicit determinant formula for the metrics come
from the following expressions for the vertex operator 
expectation values which we shall derive in the
following sections:
\debut
\vv{g} = \Om\ \cdot\ \tau(Y_j|\mu_j) \quad &;&\quad 
\vv{g^*} = \rho^{\inv{4}}\,\Om^*\ \cdot\  \tau(Y_j\,B_j|\mu_j) \label{tau1}\\
\vv{Wg} = \rho\,\Om_w\ \cdot\  \tau(Y_jB^2_j|\mu_j) \quad &;&\quad 
\vv{Wg^*} = \rho^{\inv{4}}\,\Om_w^*\ \cdot\ 
 \tau(Y_j\,B_j^{-1}|\mu_j) \label{tau2}
\fin 
The tau functions $\tau(Y|\mu)$ could be written as
$n\times n$ determinants:
\debut
\tau(Y|\mu) = \det_{n\times n}\Bigl[{ 1 + i V }\Bigr] \quad {\rm with}\quad 
V_{ij} = \frac{2 \mu_i Y_j}{\mu_i+\mu_j} \label{taudet}
\fin
The parameters are the following:
\debut
Y_j = y_j\, X_j\  \({\prod_{p=1}^m B_{jp}^{u_p} }\) \quad {\rm with} \quad 
B_{jp} = \frac{\mu_j-\mu_p}{\mu_j+\mu_p} \quad {\rm and}\quad
B_{j} = \frac{1-\mu_j}{1+\mu_j}\label{ybb}
\fin
Finally the prefactors $\Om$ and $\Om_w$ are given by:
\debut
\Om = \({\prod_{p=1}^m X_p^{u^2_p/4}}\)\ \({\prod_{p<q} B_{pq}^{u_pu_q/2}}\)
\quad &;& \quad
\Om^* = \({\prod_{p=1}^m B_p^{u_p/2}}\)\ \Om \label{defom}\\
\Om_w = \({\prod_{p=1}^m B_p^{u_p}}\)\ \Om
\quad &;& \quad
\Om^*_w = \({\prod_{p=1}^m B_p^{-u_p/2}}\)\ \Om \non
\fin
The coefficient $G_{ab}$ of the metrics are thus 
the ratio of these determinants. Note that the prefactor $\Om$
cancels when computing these ratios.

Stationary axis symmetric solutions may be obtained by
analytic continuation, see Section 6.3 below.
However the reality conditions are then more
involved and not all of these solutions are physical.
Notice also that for an infinite number of parameters
$(z_j,y_j)$ the above solutions may be described in terms of
Fredholm determinants. 

Of course many solutions of Ernst equations have already
been described in the literature, see
\cite{KhanPen,Nuktu,ChanXan,Ergarcia,letel,rerefs} for a sample of references
and \cite{Grif,Kramer} and references therein,
The formula closest to the ones we are describing here are
those obtained by P. Letelier, cf. ref.\cite{letel}. 
These later formulas also present the metric coefficients
as ratio of determinants which are
analogous to those we have described but are not quite the same.
Moreover in the formula \cite{letel} a few quadratures,
which in the formulas above are absent, remain to be solved.
But, again our principal aim was to show explicitly
how vertex operator methods, which are standard in integrable
models, may be applied to the Ernst equations.

The formulas (\ref{sig0}--\ref{g11})
for the metrics originate from an algebraic
expression for the Ernst potential which was derived in
ref.\cite{BeJu} using the algebraic method of dressing 
transformations:
\debut
\frac{\rho}{G^*_{22}} + i \frac{G^*_{12}}{G_{22}^*} =
\frac{ \vev{\Psi_0\cdot (g_-^{-1}\, W_2(1)\, g_+)\cdot \Psi_0^{-1}} }{
\vev{\Psi_0\cdot (g_-^{-1}g_+)\cdot \Psi_0^{-1}} }
\equiv  \frac{ \vv{Wg} }{\vv{g}}
\label{magic}
\fin
Here, $\Psi_0$, $W_2(1)$ and $g_\pm$ are operators acting
on an auxiliary Fock space and $\vev{\cdots}$ denotes the vacuum
expectation value in that Fock space.
$\Psi_0$ is an explicit operator defined in eq.(\ref{psivac}) below
which carries all $(z,\rho)$ dependence. $W_2(1)$ is a specific
vertex operator associated to the Kramer-Neugebauer duality
and $g=g_-^{-1}g_+$ are elements of an $SL(2,R)$ affine 
Kac-Moody group. 

However, the formula (\ref{magic}) is not the final answer since it requires
factorizing the element $g$ in two pieces as $g=g_-^{-1}g_+$.
This factorization problem is specified by Einstein's equations
up to a residual $SO(2)$ gauge freedom which reflects the gauge symmetry
of these equations. Eq.(\ref{magic}) is only true in a specific gauge:
the triangular gauge. The problem of fixing this gauge in the
factorization of the group element as $g=g_-^{-1}g_+$ 
remained unsolved in \cite{BeJu}.
The main new technical point of this paper is to
solve algebraically this gauge condition.
This will be done in three steps:
(i) by relating the triangular gauge condition to the dual
pair of solution of Ernst equations, cf Proposition 1,
(ii) by establishing the relation between pairs of group
elements $g$ and $g^*$ corresponding to dual pairs of solutions,
cf Proposition 2, and
(iii) by using these two results to factorize the element
$g=g_-^{-1}g_+$ and to compute $g_-^{-1}W_2(1)g_+$, cf Proposition 3.
This then leads to the determinant formula (\ref{tau1},\ref{tau2}).
The key algebraic result is the duality $(g,g^*)$, 
see eqs.(\ref{vertex},\ref{vertexdual}), 
or eqs.(\ref{dress1},\ref{dress2}) for its interpretation in the dressing group.

Although this paper is necessarily quite technical, we have tried to
minimize the technical aspects. In particular
the proofs could probably be omitted when reading it.
Since this paper is a continuation of ref.\cite{BeJu} and
has for aim to fill the gap left open there,
we shall freely use results from that reference.

\section{Equations of motion and dualisation.}
This section is devoted to recall
a few well known facts on  the Ernst equations:
its duality property and the relation between this property
and the triangular gauge condition.

Following Papapetrou, let us parametrize the metric (\ref{ds2}) as:
\debut
ds^2   =  \rho^{-\half } e^{2\hat \sig}~ (dz^2-d\rho^2) + \rho\De^{-1}dx^2
+ \rho \De (dy - N dx)^2 \label{ds2bis}
\fin
This means that $G_{22}=\rho \De$ and $G_{12}=-\rho \De N$.
The fields $(\hat \sig, \De,  N)$ only depends on the 
two coordinates $z$ and $\rho$.
This ansatz can be made more covariant by introducing
two light-cone coordinates $u$ and $v$ such that
\debut
\rho = a(u) + b(v) \quad, \quad z=a(u)-b(v) \label{rhozuv}
\fin
with $a$ and $b$ any functions. The metric is then:
\debut
ds^2   = - 4\rho^{-\half } e^{2\hat \sig}~ dadb + \rho\De^{-1}dx^2
+ \rho \De (dy - N dx)^2 \non
\fin
Consistency  of this parametrization follows from
the fact that Einstein equations imply that $\rho$
defined as the square root of $\det G_{ab}$ is harmonic:
$\d_u\d_v\rho=0$. In the following we shall denote
by $\d_+$ and $\d_-$ the derivatives with respect to $u$ and $v$.

As is well known, given any solution of the  reduced Einstein's 
equations one gets using the Kramer-Neugebauer symmetry
a dual solution by introducing the metric $ds_*^2$ 
parametrized as above but with the original fields 
$(\hat \sig, \De, N)$ replaced by
the dual fields $(\hat \sig^*, \De^*, N^*)$ with:
\debut
\De^* = 1/(\rho\De) \quad,\quad
\De^*\d_\nu N^*= \ep_{\nu \al} \De \d_\al N 
\quad,\quad \De^* e^{4\hat \sig^*} = \De e^{4 \hat \sig} \label{dual}
\fin
Hence, the dual metric $ds_*^2$ can be written as:
\debut
ds^2_*= \De e^{2\hat \sig} ~ (dz^2-d\rho^2)+ 
\rho^2 \De dx^2 + \De^{-1}(dy - N^* dx)^2 \label{dualbis}
\fin

\subsection{Reduced vacuum Einstein's equations.}
The vacuum Einstein's equations,
which codes for the Ricci flatness of the metric (\ref{ds2}),
reduce to the Ernst equations.
They are those of a non linear sigma model on the
coset space $SL(2,R)/SO(2)$.
They can be written using a first order formalism. 
This amounts to introduce a connection taking values 
in the $sl(2,R)$ Lie algebra but covariant with respect to 
the $so(2)$ subalgebra. We denote by $Q$ and $P$
its components with $Q$ antisymmetric  
and $P$ traceless symmetric $2\times 2$ matrices.
The equations for $Q$ and $P$ are then, 
cf eg. ref.\cite{nico} and references therein: 
\debut
[\d_+ + Q_+ + P_+ , \d_- + Q_- + P_-]&=&0 \label{EL1}\\
D_-(\rho P_+) + D_+(\rho P_-) =0 \label{EL3}\\
2(\d_\pm\hat \sig)(\rho^{-1}\d_\pm\rho)= tr(P_\pm P_\pm) \label{EL4}
\fin
where $D_\pm = \d_\pm + [Q_\pm,~\cdot~]$ is the covariant derivative. 
This system is gauge covariant with gauge group $SO(2)$. 
 It acts as 
$Q_\pm \to Q_\pm + \La^{-1}\d_\pm \La$ and $P_\pm \to \La^{-1} P_\pm \La$ 
for $\La\in SO(2)$ while $\hat \sig$ is gauge invariant.

The first equation (\ref{EL1})
means that $(Q_\pm+P_\pm)$ is flat;
this ensures that there exists a $2\times 2$ matrix $\CV$,
with $\CV\in SL(2,R)$ such that 
$\CV\d_{\pm}\CV^{-1}= Q_\pm + P_\pm$. 
The matrix $\CV$ is the matrix of the zwei-beins such that
the metric coefficients $G_{ab}$ are such that:
$$G_{ab} = \rho\ \({{}^t\CV\ \CV }\)_{ab}$$
Note that $\det G_{ab}=\rho^2$ since $\det \CV=1$.
It is often convenient to choose a gauge in which $\CV$ is triangular:
\debut
\CV= \pmatrix{\De^{-\half} & 0 \cr -N \De^\half & \De^\half \cr}
=\pmatrix{\De^{-\half} & 0 \cr 0& \De^\half \cr}
\pmatrix{1 & 0 \cr -N  & 1 \cr} \non
\fin
It reproduces the parametrization (\ref{ds2bis}) 
for the metric. In this gauge the connection is:
\debut
Q_\pm+P_\pm = \half(\De^{-1}\d_\pm\De)\cdot\sig^z 
+ \De(\d_\pm N)\cdot\sig^-     \label{Etri} 
\fin
Plugging this parametrization into eqs.(\ref{EL1},\ref{EL3}) gives
the Ernst equations.

\subsection{Alternative parametrization of Ernst equations.}
To make contact with the vertex operator approach 
described in the following we need to introduce an alternative
parametrization of the connection. Since $Q$ is antisymmetric  
and $P$ traceless symmetric, we may parametrize them as
follows:
\debut
Q_\pm &=& \half\, \({\d_\pm \phi_\pm }\)\ [\sig^+-\sig^-] \label{qpm}\\
P_\pm &=& \inv{4}(\rho^{-1}\d_\pm\rho)\, \({Z_\mp\, e^{i\phi_\mp} }\)\
\[{\sig^z-i(\sig^++\sig^-)}\] + {\rm c.c.} \non
\fin
with $\sig^z,\ \sig^\pm$ Pauli matrices. The dynamical
fields are thus the two real fields $\phi_\pm$  
and the two complex ones $Z_\pm$. 
Gauge transformations act by translations as 
$\phi_\pm\to \phi_\pm + \la$ with $\la$ a real function. 
In particular $Z_\pm$ as well as $\phi=\phi_+-\phi_-$ are
gauge invariant.

In this parametrization, the Ernst equations become:
\debut
\d_\pm Z_\pm &=& \inv{2}(\rho^{-1}\d_\pm\rho)\,
\({Z_\pm - e^{\mp i\phi}\, Z_\mp}\) \label{ddphi}\\
\d_+\d_-\phi&=& -\frac{i}{2}(\rho^{-1}\d_+\rho)(\rho^{-1}\d_-\rho)\
\({ Z_+\bar {Z_-} \, e^{i\phi} - \bar {Z_+}\, Z_-\, e^{-i\phi} }\) 
\label{ddphi2}\\
4 \d_\pm \hat \sig &=& (\rho^{-1}\d_\pm\rho)\, Z_\mp \bar {Z_\mp} 
\label{ddphi3}
\fin
These equations are gauge invariant since 
$\phi$ and $Z_\pm$ are gauge invariant.
The equivalence between eqs.(\ref{ddphi}) and the more usual Ernst
equations follows by plugging the parametrization (\ref{qpm}) in 
the first order formalism eqs.(\ref{EL1},\ref{EL3}).
Note the similarity with the sine-Gordon equation.

Let us also mention that eqs.(\ref{ddphi},\ref{ddphi2},\ref{ddphi3})
may be rewritten in a bilinear form similar to Hirota's equations.
Namely, let $\tau_0$ and $\tau_\pm$ be tau-functions defined by:
\debut
\tau_0=\exp({\hat \sig-\frac{i}{4}\phi}),\quad
\tau_+= \bar {Z_+}\ \tau_0, \quad
\tau_-= Z_-\ \tau_0 \non
\fin
then, eqs.(\ref{ddphi},\ref{ddphi2},\ref{ddphi3}) are equivalent to
the following bilinear equations:
\debut
\tau_0\, (\d_\pm\tau_\pm)- \tau_\pm\, (\d_\pm\tau_0) 
&=& \half(\rho^{-1}\d_\pm\rho)\, 
(\tau_0\tau_\pm - \bar \tau_0 \bar \tau_\mp) \label{hir}\\
\tau_0\, (\d_+\d_-\tau_0) - (\d_-\tau_0)\, (\d_+\tau_0) 
&=& -\inv{4}(\rho^{-1}\d_+\rho)(\rho^{-1}\d_-\rho)\, 
\bar \tau_+ \, \bar \tau_- \non
\fin
This is proved by direct substitution.
Eqs.(\ref{hir}) are Hirota's form of Ernst's equations.

\subsection{Dualisation and triangular gauge.}
Let us now explain the interplay between the 
condition for choosing the triangular gauge and
the existence of two dual solutions.

\proclaim Proposition 1.
Let $(\phi_\pm, Z_\pm)$ and $(\phi_\pm^*,Z_\pm^*)$ be the
components of the two dual solutions assuming that 
both connections are in the triangular gauge.
The duality relation is then equivalent to:\\
--- a duality relation between the fields $\phi_\pm$ and $\phi_\pm^*$, 
which is valid only in the triangular gauge: 
\debut
\phi_\pm = \half \({ \phi^* \pm \phi }\) =\pm \phi_\pm^* \label{phistar},
\fin
--- a duality relation between $Z_\pm$ and $Z^*_\pm$, which is gauge invariant:
\debut
Z^*_+ + Z_+ &=& - e^{-i(\phi+\phi^*)/2} \label{zzstar}\\
Z^*_- + \bar {Z_-} &=& - e^{-i(\phi-\phi^*)/2} \non
\fin

\proof
Indeed, let $(\phi_\pm, Z_\pm)$ be a 
solution of the Ernst equations (\ref{ddphi},\ref{ddphi2}).
Imposing the triangular gauge as in eq.(\ref{Etri})
demands that the coefficient of $Q_\pm+P_\pm$ along $\sig^+$ be zero.
In the parametrization (\ref{qpm}) this amounts to impose:
\debut
2\,\d_\pm \phi_\pm = i (\rho^{-1}\d_\pm\rho)\,
\({ Z_\mp\, e^{i\phi_\mp} - \bar {Z_\mp}\, e^{-i\phi_\mp} }\)
\label{triang}
\fin
Let us now define the dual solution $\phi^*$ and $Z^*_\pm$ by
the formulas (\ref{phistar}) and (\ref{zzstar}). 
Notice that when parametrized in
terms of $\phi$ and $\phi^*$, the condition (\ref{triang})
becomes:
\debut
\d_\pm\({ \phi^*\pm\phi}\)
= i(\rho^{-1}\d_\pm\rho)\, \[{ Z_\mp\, e^{\frac{i}{2}(\phi^*\mp\phi)}
- \bar {Z_\mp}\, e^{-\frac{i}{2}(\phi^*\mp\phi)}  }\] \non
\fin
One has to prove that $(\phi^*,Z^*_\pm)$
are solutions of eqs.(\ref{ddphi},\ref{ddphi2}).
Consider eq.(\ref{ddphi}) for $Z_+^*$.
Using the definition of $Z^*_+$ and eq.(\ref{ddphi}) satisfied by $Z_+$
and eq.(\ref{triang}) for $\phi+\phi^*$, we get:
\debut
\d_+ Z^*_+ &=& -\half (\rho^{-1}\d_+\rho)\,
\[{ Z_+ - e^{-i\phi}Z_- + \({ Z_-e^{\frac{i}{2}(\phi^*-\phi)}
- \bar {Z_-}e^{-\frac{i}{2}(\phi^*-\phi)} }\)
e^{-\frac{i}{2}(\phi^*+\phi)}  }\]\non\\
&=& \half (\rho^{-1}\d_+\rho)\,
\[{Z_+^* - Z_-^* e^{-i\phi^*} }\] \non
\fin
This shows that $Z_+^*$ is the solution. The equation for
$Z_-^*$ is proved in a similar way. The equation
for the dual field $\phi^*$ is proved by taking derivatives 
of eq.(\ref{triang}).
Finally, since the transformation (\ref{zzstar}) is an involution
the triangular gauge condition for $\phi^*_\pm$ follows from
that of $\phi$, eq.(\ref{triang}).
\square

Remark that the duality relation (\ref{phistar}) 
and (\ref{zzstar}) are purely algebraic and that 
eq.(\ref{phistar}) is similar to the usual
abelian T-duality of string theory.
This proposition also shows that solving for the
triangular gauge or for the dual solution are 
equivalent problems.

Using the defining relation between $\phi,Z_\pm$ and
the connection $Q_\pm + P_\pm $ in the triangular gauge,
one may translate the duality (\ref{phistar},\ref{zzstar}) 
on the connection as:
\debut
Q_\pm+P_\pm &=& \half(\De^{-1}\d_\pm\De)\cdot\sig^z + \De(\d_\pm N)\cdot\sig^-
\label{pqdual}\\
\to\qquad 
Q^*_\pm+P^*_\pm &=& -\half(\rho^{-1}\d_\pm\rho)\cdot\sig^z
-\half(\De^{-1}\d_\pm\De)\cdot\sig^z \pm \De(\d_\pm N)\cdot\sig^-\non 
\fin
This is clearly equivalent to the Kramer-Neugebauer duality (\ref{dual}).

\section{Integrability and vertex operators.}
The aim of this section is to recall the use of
vertex operators for solving the Ernst equations 
following the method initiated in ref.\cite{BeJu}.
This method was based on an application of the 
dressing method \cite{sem}. We have written this section
in a down-to-earth way by just presenting propositions
which give the rules for computing solutions of
the Ernst equation using vertex operators. These rules
are analogous of the famous Kyoto formula \cite {kyoto} for the tau function
of the KP hierarchy in terms of fermions or vertex operators.
The logic which allows us to go from the dressing method
to these vertex operator formulas is recalled in Appendix A.
We shall also describe the dual pairs of vertex operators
corresponding to dual pairs of solutions

\subsection{Vertex operators and tau functions.}
Vertex operators are exponentials of a free bosonic field acting 
on an auxiliary Fock space. They may be used to find solutions
of the Ernst equations in a way similar to their use for solving
the KP hierarchy as described the Kyoto school \cite {kyoto}.
Let us denote by $\hat X(w)$ the bosonic  field:
\debut
\hat X(w)=-i\sum_{n~odd}~p_{-n}\frac{w^n}{n}\quad {\rm with}\quad
[p_n,p_m]=n\de_{n+m,0} \non
\fin
The operators $p_n$ generate a Fock space,
we denote by $\ket{0}$ its vacuum: $p_n\ket{0}=0$ for $n>0$.
For any number $u$, let $W_u(w)$ be the vertex operators:
\debut
W_u(w)= :\exp(-iu \hat X(w)): \label{verop}
\fin
The double dots refer to the normal ordering
which amounts to move to the right the oscillators $p_n$
with $n$ positive.
The parameter $w$ is called the spectral parameter.
The Virasoro algebra acts on the Fock space generated by the $p_n$.
The Virasoro generators $L_n$ are represented by
$$
\sum_n (L_n-\inv{16}\de_{n,0})w^{-2n-2} = -\inv{4}: (\d_w \hat X)^2:
$$

The algebraic dressing method applied to the Ernst equations 
leads to the following result:

\proclaim Proposition \cite{BeJu}.
Let $\Psi_0$ be defined as:
\debut
\Psi_0 = \({ \frac{\rho+z+1}{2\rho} }\)^{L_0-L_1} \
\({ \frac{\rho+z+1}{2} }\)^{L_0-L_{-1}} \label{psivac}
\fin
Let $g$ be any product of vertex operators of the following form:
\debut
g={\rm const.}\ \prod_{p=1}^mW_{u_p}(w_p)\cdot 
\prod_{j=1}^n\({ 1 + i y_j W_2(w_j) }\) \label{gvertex}
\fin
where $(y_j,w_j)$ and $(u_p,w_p)$ are $2(m+n)$ real parameters,
then the fields $(\hat \sig,\phi, Z_\pm)$,
defined by the following expectation values,\\
\debut
\exp\({\hat \sig - \frac{i}{4}\phi }\) &=& 
\vev{\Psi_0 g \Psi_0^{-1}} \equiv \vv{g}\label{tau0}\\
\bar {Z_+}\ \exp\({\hat \sig - \frac{i}{4}\phi }\) &=&
\vev{(\Psi_0 g \Psi_0^{-1})\cdot p_{-1}} \equiv \vv{gp_{-1}}
\label{tau+}\\
{Z_-}\  \exp\({\hat \sig - \frac{i}{4}\phi }\) &=& -
\vev{p_1\cdot(\Psi_0 g \Psi_0^{-1})} \equiv -\vv{p_1 g}
\label{tau-}
\fin
are solutions of equations (\ref{ddphi},\ref{ddphi2},\ref{ddphi3}).

The above equations serve as the definition of 
$\vv{g}$, $\vv{gp_{-1}}$ and $\vv{p_1 g}$.
This is the notation used in Section 2,  eqs.(\ref{sig0}--\ref{g11}).
These expectation values are the tau-functions of the model.

To make sense the vertex operators in eq.(\ref{gvertex}) have to be
ordered in decreasing order of the $|w_j|$'s. Modification
of this order may be done by analytic continuation in
the spectral parameters.

To compute these expectation values one needs 
to know how to conjugate vertex operators $W_u(w)$ with $\Psi_0$.  
One has \cite{BeJu}:
\debut
\Psi_0\cdot W_u(w_j)\cdot \Psi_0^{-1} = X_j^{u^2/4}\cdot
 W_u\({\mu_j}\)\cdot \label{wmuAB}
\fin
with $X_j$ and $\mu_j$ defined in eq.(\ref{defY}) with the
parameter $z_j=\frac{w_j^2+1}{w_j^2-1}$.
Notice in particular that $\Psi_0\cdot W_u(1)\cdot \Psi_0^{-1}
= \rho^{\frac{u^2}{4}}~ W_u(1)$.
With this result in hand, the computation of the 
expectation values (\ref{tau0},\ref{tau+},\ref{tau-})
is reduced to the computation of expectation values of vertex operators.
As recalled in Appendix B, this is done using the usual
Wick's theorem. For example:
\debut
\vv{g} &=& (\prod_{p=1}^m X_p^{u^2_p/4})\ 
\vev{\prod_{p=1}^mW_{u_p}(\mu_p)\cdot 
\prod_{j=1}^n\({ 1 + i Y_j W_2(\mu_j) }\) }  \non\\
&=& \Om\, \cdot \, \tau(Y_j|\mu_j) \label{vvg}
\fin
with $\Om$ and $\tau(Y|\mu)$ defined in 
eqs.(\ref{taudet},\ref{defom}) above. 
The expectation values $\vv{gp_{-1}}$ and $\vv{p_1g}$ are computed 
similarly using the formula recalled in Appendix B. 
They may be obtained from the previous expression 
for $\vv{g}$ by replacing each monome $X_{k_1}\cdots X_{k_p}$
by,
\debut
X_{k_1}\cdots X_{k_p} &\to&
X_{k_1}\cdots X_{k_p}\ \({\sum_m u_{k_m}\mu_{k_m}^{-1} }\) 
\quad {\rm for}\quad \vv{gp_{-1}} \label{vgp}\\
X_{k_1}\cdots X_{k_p} &\to&
-\ X_{k_1}\cdots X_{k_p}\ \({\sum_m u_{k_m}\mu_{k_m} }\) 
\quad {\rm for}\quad \vv{p_1g} \non
\fin
in the formula (\ref{vvg}).

\subsection{Vertex operators and duality.}

Two dual solutions correspond to two dual products of
vertex operators that we shall denote by $g$ and $g^*$.  
The duality relations (\ref{phistar}) and (\ref{zzstar})
can be translated into quadratic relations for
the expectation values of these operators similar to the
Hirota equations. Namely:
\debut
\vv{g}\ \vv{g^*p_{-1}} + \vv{g^*}\ \vv{gp_{-1}}
&=& - \bar {\vv{g}}\  \bar {\vv{g^*}} \label{ladu1}\\
\vv{g}\ \bar{\vv{p_1g^*}} + \bar {\vv{g^*}}\ \vv{p_1g}
&=& + \bar {\vv{g}}\ \vv{g^*} \label{ladu2}
\fin

\proclaim Proposition 2.
Pairs of solutions of the above duality relations,
eqs.(\ref{ladu1},\ref{ladu2}), are provided by the
following pairs of vertex operators:
\debut
g&=& {\rm const.}\ \prod_{p=1}^mW_{u_p}(w_p)\cdot 
\prod_{j=1}^n\({ 1 +i y_j W_2(w_j) }\) \label{vertex}\\
g^* &=&{\rm const.}\  W_{-1}(1)\cdot \prod_{p=1}^mW_{-u_p}(w_p)\cdot 
\prod_{j=1}^n\({ 1 + i y_j W_{-2}(w_j) }\)  \label{vertexdual}
\fin
Up the multiplication by $W_{-1}(1)$ the dual
vertex operator is obtained by the charge $u$ into $-u$.
The constant prefactors in eqs.(\ref{vertex},\ref{vertexdual})
are irrelevant.

\proof
It relies on an identity for the tau-functions
proved in ref.\cite{BaBe}.
Let us sketch the proof of eq.(\ref{ladu1}) for $m=0$.
Recall eq.(\ref{vvg}) for $\vv{g}$ and eq.(\ref{vvgstar}) 
below for $\vv{g^*}$:
\debut
\vv{g} = \tau(Y_j|\mu_j) \quad {\rm and}\quad 
\vv{g^*} = \rho^{\inv{4}}\,\tau(B_j\,Y_j|\mu_j)
\fin
From eqs.(\ref{vgp}) one infers  that the expectation values
$\vv{gp_{-1}}$ and $\vv{g^*p_{-1}}$ may be written in terms
of derivatives of tau-functions. Namely:
\debut
\vv{gp_{-1}} &=&\frac{\d}{\d\mu_0}\tau(B_{0j}Y_j|\mu_j)\vert_{\mu_0=0} \non\\
\rho^{-\inv{4}}\,\vv{g^*p_{-1}} &=&-\tau(B_j\,Y_j|\mu_j)-
\frac{\d}{\d\mu_0}\tau(B_{0j}B_jY_j|\mu_j)\vert_{\mu_0=0} \non
\fin
with $B_{0j}=\frac{\mu_0+\mu_j}{\mu_0-\mu_j}$.
Eq.(\ref{ladu1}) may then be written as a bilinear
identity for the tau-functions. The later follows 
by adding and expending in power of $\mu_0$ the
following two relations proved in \cite{BaBe}:
\begin{eqnarray}
\tau(B_{0k}B_{1k} Y_k) \bar {\tau(Y_k)} +
\bar {\tau(B_{0k}B_{1k} Y_k)} \tau(Y_k) 
&=& \tau(B_{0k} Y_k) \bar {\tau(B_{1k}Y_k)} +
\bar {\tau(B_{0k} Y_k)} \tau(B_{1k}Y_k) \label{EBi}\\
B_{10}\Big[\tau(B_{0k}B_{1k} Y_k) \bar {\tau(Y_k)} -
\bar {\tau(B_{0k}B_{1k} Y_k)} \tau(Y_k) \Big]
&=& \tau(B_{0k} Y_k) \bar {\tau(B_{1k}Y_k)} -
\bar {\tau(B_{0k} Y_k)} \tau(B_{1k}Y_k) \label{EBii}
\end{eqnarray}
with $B_{1k}=B_k$ and $B_{10}=\frac{1-\mu_0}{1+\mu_0}$.
The proofs of the general case $m\not= 0$ as well as eq.(\ref{ladu2})
are similar.
\square

In eq.(\ref{vertexdual}) the order of the operators matters:
the operator $W_{-1}(1)$ has to be on the left.
When changing the order of the operators one has to
take into account their commutation relations,
ie. $W_{-1}(1)$ anticommutes with $W_{-2}(w)$.

Expectation values of the dual vertex operators may be
evaluated using the conjugation formula (\ref{wmuAB}) and
Wick's theroem as explained for $\vv{g}$. One gets:
\debut
\vv{g^*} &\equiv& \vev{\Psi_0\, g^*\, \Psi_0^{-1}} \non\\
&=& \rho^{\inv{4}}\,\Om^*\ \cdot\  \tau(B_j\,Y_j|\mu_j) \label{vvgstar}
\fin
as in eq.(\ref{tau1}).

\section{Algebraic computation of the metric coefficients.}

The previous section made precise the relation between 
the two dual vertex operators $g$ and $g^*$. 
It allows us to compute algebraically  the 
gauge invariant fields $(\hat \sig,\phi, Z_\pm)$ and their
duals $(\hat \sig^*,\phi^*, Z_\pm^*)$.
This is not quite the final answer for the metric
since to go from $(\phi,Z_\pm)$ to the metric coefficients $G_{ab}$,
or $\De$ and $N$, one needs to impose the triangular
gauge and then integrate the connection $Q_\pm+P_\pm$ 
to obtain the zwei-bein $\CV$. Imposing the triangular gauge
requires solving eq.(\ref{triang}). This is a non-linear
problem which was actually not solved in \cite{BeJu}.
We shall now solve it using our knowledge on the dual pairs
of vertex operators, eqs.(\ref{vertex},\ref{vertexdual}), and 
on the link between the duality and the triangular gauge,
eqs.(\ref{phistar},\ref{zzstar}).
We will then be able to use formula (\ref{magic}) to compute
the metric coefficients $G_{ab}$.

\subsection{Vertex operator representation and factorization.}
In order to be able to apply formula (\ref{magic}) we need
to make a small detour into group theory in order to
explain the relation between factorization problem
in affine Kac-Moody group and vertex operators.

This relation arises because the vertex operators (\ref{verop})
may be used to represent the $sl(2,R)$ affine algebra
on the Fock space. The  commutation relations of
the $sl(2,R)$ affine Kac-Moody algebra are:
\debut
\[{ x\otimes t^n, y\otimes t^m }\]&=&
[x,y]\otimes t^{n+m} +n \frac{k}{2} ~tr(xy)\de_{n+m,0} \label{comh}
\fin 
The affine Kac-Moody $sl(2,R)$ algebra is twisted in the sense
that its elements $x\otimes t^n$ are such that $x\in so(2)$
if $n$ is even while $x$ is an $2\times 2$ traceless symmetric
matrix if $n$ is odd.
We are actually considering the semi direct product
of the Virasoro algebra with the affine algebra. The
crossed Lie bracket is:
$\[{L_n, x\otimes t^m}\] = -\frac{m}{2} x\otimes t^{m+2n}$.

The representation of the affine algebra on the Fock space
is specified by the following relations \cite{lewi}:
\debut
iw \frac{d\hat X(w)}{dw} &=& \sum_{n~odd}
(\sig^z\otimes t^n) w^{-n} \label{repv}\\
i~ W_2(w) &=& 2 \sum_{n~even}
((\sig^+-\sig^-)\otimes t^n)w^{-n}
-2\sum_{n~odd}((\sig^++\sig^-)\otimes t^n)w^{-n} \non
\fin
This is a highest weight representation.
The highest weight vector is identified with the
vacuum vector $\ket{0}$. It is such that
$(\sig^+-\sig^-)\ket{0}= \frac{i}{2}\ket{0}$
and $(\sig^+-\sig^-)\otimes t^n\ket{0}=0$ for $n>0$.
Remark that in particular eq.(\ref{repv}) means that 
$(\sig^z\otimes t^n)$ is represented by the bosonic
oscillator $p_n$. 

The algebraic dressing method from which equation (\ref{magic})
follows relies on a factorization problem specified on
the affine $SL(2,R)$ Kac-Moody group. This is defined
as follows. Let $\CB_\pm$ be the two Borel
subalgebras respectively generated by the central charge $k$
and the elements $x\otimes t^{\pm n}$ with $n$ positive.
Let $B_\pm=\exp\,\CB_\pm$ the corresponding Borel subgroups.
Then the factorization amounts to formally decompose any element
$g$ in the affine $SL(2,R)$ Kac-Moody group 
as the product of elements in $B_\pm$:
\debut
g= g_-^{-1}g_+ \quad {\rm with}\quad  g_\pm \in B_\pm \label{facto}
\fin
where we also demand that the components of $g_\pm$ on the
exponential of the central charge is inverse.
Remark that since the two Borel subalgebra $\CB_\pm$
have elements $x\otimes t^0$ in common, the above
factorization is defined only up to a multiplication
$g_\pm\to h g_\pm$ by elements $h\in SO(2)$.
As explained in ref.\cite{BeJu}, this freedom  is
linked to the $SO(2)$ gauge symmetry of Ernst 
equations (\ref{EL1},\ref{EL3}).

We can now state a result from \cite{BeJu}:

\proclaim Proposition \cite{BeJu}.
Let $\hat \CE= \hat \CE_+(1)-\hat \CE_-(1)$,
let $g=g_-^{-1}g_+$ be factorized according to the triangular gauge,
then
\debut
\frac{1}{\De^*} \ - i N^* =
-i\frac{\vev{\Psi_0\cdot (g_-^{-1} \hat \CE g_+)\cdot \Psi_0^{-1}}}{
\vev{\Psi_0\cdot (g_-^{-1}g_+)\cdot \Psi_0^{-1}}}
\label{magic3}
\fin
Furthermore, in the Fock space the elements $\hat \CE$ 
is represented: $\hat \CE = i W_2(1)$.\\
A similar expression holds with the dual operator $g^*$ and $\De$ and $N$.

Eq.(\ref{magic3}) is only valid in the triangular gauge.
Let us explain in more detail how this gauge choice fixes
the factorization of $g$ as $g_-^{-1}g_+$. 
Since $g_\pm\in B_\pm$, one has:
\debut
g_\pm = \exp\(\pm \frac{\zeta}{2} k\)\  
\exp\(-\frac{\varphi_\pm}{2}(\sig^+-\sig^-)\)\ 
\times \[|{\rm degree}|\ \geq 1 \] \non
\fin
For a given group element $g=g_-^{-1}g_+$ only 
the difference $\varphi_+ -\varphi_-$ is fixed.
To translate $\varphi_\pm$ keeping this difference fixed
amounts to multiply $g_\pm\to hg_\pm$ with $h\in SO(2)$.
The link between this freedom and the $SO(2)$ gauge freedom
of the Ernst equation relies on the fact \cite{BeJu} that
$\varphi_\pm$ coincide with the fields $\phi_\pm$ at $z=0,\ \rho=1$.
Thus in the triangular gauge, 
\debut
\varphi_\pm = \half (\varphi^* \pm \varphi ) \label{varp}
\fin
with $\varphi$ and $\varphi^*$ equal to $\Phi$ and $\Phi^*$ at $z=0,\rho=1$.
This follows from eq.(\ref{phistar}). 
Since $\Psi_0=1$ at $z=0,\ \rho=1$, they may be evaluated
using eq.(\ref{tau0}):
\debut
\exp\({i\frac{\varphi}{2}}\) = \frac{\bar {\vev{g}}}{\vev{g}}
\quad ,\quad
\exp\({i\frac{\varphi^*}{2}}\) = \frac{\bar {\vev{g^*}}}{\vev{g^*}}  \label{varstar}
\fin
Here, the expectation values are the vacuum expectation values, without
insertions of $\Psi_0$.

\subsection{Factorization and dualisation.}
The factorization problem in the affine Kac-Moody group may be
understood as a kind of normal ordering.
So  when considering the vertex operator representation
one has to face two different normal orderings: the one associated
to bosonic oscillators $p_n$ and the group theoretical one.
We shall now explain the link between them for the
vertex operators (\ref{vertex}).

First, consider vertex operators $W_u(w)$. 
Since $(\sig^z\otimes t^n)$ is represented by $p_n$,
we may consider them as elements of the affine group.
Namely:
\debut
W_u(w)&=& 
\exp\({-\frac{u}{2} \sig^z\otimes\log\({\frac{1+w/t}{1-w/t}}\)}\)\cdot
\exp\({\frac{u}{2} \sig^z\otimes\log\({\frac{1+t/w}{1-t/w}}\)}\) \label{wuh}\\
&\equiv& W_u(w)_-^{-1} \cdot W_u(w)_+ \non
\fin
The last equation serves as definition of $W_u(w)_\pm$, elements
of the Borel subgroups $B_\pm$. Thus, the two normal 
orderings coincide for these group elements.

Consider now the product of vertex operators of the form
$\prod_j\({ 1 +i y_j W_2(w_j)}\)$.
Since $W_2(w)$, which are generating functions representing
elements of the affine algebra, are nilpotent 
inside any correlation functions, ie. $W_2(w)W_2(w)=0$, 
these products are representations of elements of the Kac-Moody group.
As shown in ref.\cite{BaBe} these products may be factorized in the 
affine Kac-Moody group.
More precisely, let $g_\pm(j)$ be the elements of 
the Borel subgroups $B_\pm$ defined by
\debut
g_\pm(j)= \exp\(\pm \frac{r_j}{2} k\)\ 
\exp\(\frac{v_j}{2}(\sig^+-\sig^-)\)\ 
\exp\(\frac{s_j}{2} \hat \CE_\pm(w_j)\) \label{defg-g+}
\fin
with
\debut
\hat \CE_\pm(w)= \pm \[{
(\sig^+-\sig^-)\otimes\({\frac{1+(t/w)^{\pm2}}{1-(t/w)^{\pm2}}}\)
-(\sig^++\sig^-)\otimes\({\frac{2(t/w)^{\pm1}}{1-(t/w)^{\pm2}}}\) }\]
\label{defxmu}
\fin
Then, for $k=1,\cdots,n$ one has \cite{BaBe},
\debut
g_-^{-1}(1)\cdots g_-^{-1}(k)\cdot g_+(k)\cdots g_+(1)=
\prod_{j=1}^k\({ 1 +i  y_j W_2(w_j)}\) \label{g-g+w}
\fin
Eq.(\ref{g-g+w}) is valid in the Fock space representation.
The relation between the parameters $(s_j,r_j,v_j)$ and $(y_j,w_j)$
is explained in the following proposition.

We can then solve for the factorization problem:

\proclaim Proposition 3.
For $g$ and $g^*$ the dual vertex operators (\ref{vertex}) and
(\ref{vertexdual}), then:
\debut
Wg &\equiv& g^{-1}_- W_2(1) g_+ = \ (-ia+bW_2(1))\cdot
\prod_{p=1}^mW_{u_p}(w_p)\cdot 
\prod_{j=1}^n\({ 1 +i y_j W_2(w_j) }\) \label{wg}\\
Wg^* &\equiv& g^{*\, -1}_- W_2(1) g^*_+ 
= \ 
(-ia^*+b^*W_1(1))\cdot \prod_{p=1}^mW_{-u_p}(w_p)\cdot 
\prod_{j=1}^n\({ 1 + i y_j W_{-2}(w_j) }\) \label{wgstar}
\fin
in the triangular gauge.
Since the Ernst potential is defined up to a multiplicative
 constant and a constant translation on $N$,
 $a$ and $b$ are irrelevant when computing the metrics.

\proof
This relies on the relation \cite{BaBe} between 
the parameters $(y_j,w_j)$ involved in the vertex operators
and the parameters $(s_j,r_j,v_j)$ in the group elements $g_\pm(j)$
such that:
\debut
g_-^{-1}(1)\cdots g_-^{-1}(n)\cdot g_+(n)\cdots g_+(1)=
\prod_{j=1}^n\({ 1 +i  y_j W_2(w_j)}\) \label{g+n}
\fin
Let us introduce the obvious notation
$\varphi_j$ and $\varphi^*_j$ for $j\leq n$ by
\debut
\varphi_j = - 2 \sum_{k=1}^j s_k \quad,\ \quad
\varphi_j^* = - 2 \sum_{k=1}^j v_k \non
\fin
such that $s_j = -\half(\varphi_j - \varphi_{j-1})$ and similarly for $v_j$.
Of course $\varphi_\pm = \half(\varphi^*_n\pm \varphi_n)$.
First $(s_j,r_j)$ are recursively determined as functions of $(y_k,w_k)$
with $k\leq j$ by computing the expectation values of eq.(\ref{g-g+w}):
\debut
\exp\({ \sum_{k=1}^j(r_k+ i\frac{s_k}{2})}\) = 
\vev{\prod_{k=1}^j(1 + iy_k W_2(w_k))} \non
\fin
These equations determine $\varphi_j(y_k)$ as functions of $y_k$,
$k\leq j$. The $v_j$'s are then given by \cite{BaBe}:
\debut
\varphi^*_j(y_k) = \varphi_j(\beta_{j+1,k}y_k) 
\quad {\rm with}\quad \beta_{j+1,k} = \frac{w_{j+1}-w_k}{w_{j+1}+w_k} \non
\fin
This leaves $v_n$, which actually cancels in eq.(\ref{g+n}), undertermined.
However, $v_n$, or equivalently $\varphi_n^*$,
is fixed once we impose the triangular gauge.
Indeed the triangular gauge condition written as in eq.(\ref{varstar})
and formulas (\ref{vertex},\ref{vertexdual}) for the dual vertex
operators leads to:
\debut
\varphi^*_n(y_j) = \varphi_n(\beta_j y_j) 
\quad {\rm with}\quad \beta_{j} = \frac{1-w_j}{1+w_j} \label{extra}
\fin
This allows us to go one step further in the
recursion relation (\ref{g-g+w}) by inserting one
extra vertex operator with spectral parameter $w$ equals
to $1$. Let $g_\pm(n+1)$ be such that
$g^{-1}_-(n+1)g_+(n+1) = \hat a + i \hat b W_2(1)$ with $\hat a,\hat b$
 numbers.
The triangular gauge condition (\ref{extra}) then
implies:
\debut
g_-^{-1}(1)\cdots g_-^{-1}(n)\cdot (\hat a + i \hat b W_2(1))\cdot g_+(n)\cdots g_+(1)=
(a + i b W_2(1))\cdot \prod_{j=1}^n\({ 1 +i  y_j W_2(w_j)}\) \non
\fin
with $a,b$ functions of $\hat a, \hat b$. Taking $\hat a=0$ and $\hat b=1$, 
this proves eq.(\ref{wg}).
The dual equation (\ref{wgstar}) is proved similarly. 
\square

Once the operators $g^{-1}_- W_2(1) g_+$ and $g^{*\, -1}_- W_2(1) g^*_+$
are expressed in terms of the product of vertex operators, it
easy to conjugate them with $\Psi_0$ using eq.(\ref{wmuAB}).
One may then evaluate $\vv{Wg}$ and $\vv{Wg^*}$ using Wick's theorem,
cf. Appendix B. Of course one gets eq.(\ref{tau2}).
This ends the algebraic proof of the determinant formula for the
metrics.

Remark that eq.(\ref{EBi}) for the tau-functions
with $\mu_0=1$ implies:
\debut
{\rm Re}\({ \vv{g} \bar{\vv{Wg}} }\) = \sqrt{\rho}\ | \vv{g^*} |^2 \non
\fin
This shows that $G^*_{ab}$ defined in eqs.(\ref{g22}--\ref{g11})
satisfies $\det G^*_{ab}=\rho^2$.  
It provides a non-trivial check of the construction.

\subsection{Dualisation in the dressing group.}
The dressing group is the group whose
elements are the pairs $(g_-,g_+)$ factorizing $g$ as $g=g_-^{-1}g_+$.
It is different from the affine $SL(2,R)$ Kac-Moody group
since their multiplication laws do not coincide, \cite{sem}. 
In the dressing group the product is given by 
$(g_-,g_+)(h_-,h_+)=(g_-h_-,g_+h_+)$.

The solutions we have obtained should actually be labeled by 
elements of the dressing group since this is the solution generating group.
As a consequence, the duality between the vertex operators 
(\ref{vertex}) and (\ref{vertexdual}) should be 
thought of as a duality in the dressing group.
Writing the vertex operator (\ref{vertex}) in the dressing
group amounts factorizing them according to the rules
explained in the previous section:
\debut
g^{-1}_- g_+ = \prod_{p=1}^mW_{u_p}(w_p)_-^{-1}\cdot 
\hat g_-^{-1}(1)\cdots \hat g_-^{-1}(n)\cdot \hat g_+(n)\cdots \hat g_+(1)
\cdot \prod_{p=1}^mW_{u_p}(w_p)_+
\label{dress1}
\fin
where the middle term corresponds the
factorization of $\prod_{j=1}^n\({ 1 +i\, \hat  y_j\, W_2(w_j)}\)$
according to eq.(\ref{g-g+w}),
\debut
\hat g_-^{-1}(1)\cdots \hat g_-^{-1}(n)\cdot \hat g_+(n)\cdots \hat g_+(1)=
\prod_{j=1}^n\({ 1 +i\, \hat  y_j\, W_2(w_j)}\) \non
\fin
Here $\hat y_j = y_j \prod_p\beta_{jp}^{u_p}$.
Similarly the dual vertex operator is factorized as:
\debut
g^{*\, -1}_- g^*_+ &=&
 W_{-1}(1)_-^{-1}\cdot \prod_{p=1}^mW_{-u_p}(w_p)_-^{-1}\cdot 
\hat g_-^{*\, -1}(1)\cdots \hat g_-^{*\, -1}(n) \cdot \non\\
&&\cdot\hat g_+^*(n)\cdots \hat g_+^*(1) 
\cdot \prod_{p=1}^mW_{-u_p}(w_p)_+\cdot W_{-1}(1)_+   \label{dress2}
\fin
where
\debut
\hat g_-^{*\, -1}(1)\cdots \hat g_-^{*\, -1}(n)
\cdot \hat g_+^*(n)\cdots \hat g_+^*(1)=
\prod_{j=1}^n\({ 1 +i\, \hat  y_j^*\, W_{-2}(w_j)}\) \non
\fin
with $\hat y_j^* = \beta_j\, \hat y_j$. 

In eqs.(\ref{dress1},\ref{dress2}) the dual elements $(g_-,g_+)$
and $(g^*_- ,g^*_+)$ are written as elements of the dressing group
(and not in a particular representation). It is then clear that
the map from $(g_-,g_+)$ to $(g^*_- ,g^*_+)$ is an involution.
However it is not a group automorphism 
\footnote{Note however that the relation between $g^*$ and $g$ may be written
as $g^*=W_{-1}(1) \CT (g)$ with $\CT$ the automorphism of $sl(2,R)$ fixing 
$so(2)$ and multiplying $2\times 2$ traceless symmetric matrices by minus one.}.
It is unfortunate and frustrating that 
we do not know how to write this involution in more group
theoretical way without relying on these particular elements.
A better group theoretical understanding of the dualisation,
ie. of the relation between $g$ and $g^*$, will provide a way
to decipher how general the duality property and the involution
trick are and whether they apply to other integrable systems.

\section{A few examples.}
Since our aim was to describe the use of vertex operators for
solving Ernst equations and not to describe the physical
properties of the solutions, we shall only discuss 
a few examples (which are actually already known 
in the literature, cf eg. \cite{KhanPen,Nuktu,ChanXan,Ergarcia}
and \cite{Grif} and references therein).

\def\bacs{\backslash}

\subsection{Diagonal solutions.}
Diagonal solutions correspond to solutions for which $G_{ab}$ is diagonal,
ie. $N=0$. They are obtained by imposing $y_j=0$ in the parametrization 
of Section 2. They correspond to vertex operators of the form
\debut
g^{-1}_- g_+ = \prod_{p=1}^m\, W_{u_p}(w_p) \label{diag}
\fin
They depend on the $2m$ parameters $(z_p,u_p)$; 
recal that $z_p=\frac{w_p^2+1}{w_p^2-1}$.
Here and in the following examples, we drop insignificant multiplicative 
constants in front of the vertex operators.

The simplest of such solutions is the well 
known Khan-Penrose metric \cite{KhanPen}. 
It describes the interaction region of two plane impulsive gravitational waves 
having their polarization vectors aligned. 
For more details on this subject, see eg ref.\cite{Grif}.
It corresponds to the following set of parameters,
$\{(z_p,u_p)\}=\{(1,1),(-1,1)\}$, 
or equivalently to the following vertex operator: 
$$W_{1}(\infty)W_{1}(0)$$  
Notice that this choice of singular values for $z_p$ leads to null values for $X_p$.
However, we may reabsorb these  singular constant factors in the normalisation 
since Einstein's equations determine the conformal factor only up to an additive constant.  
In order to have the formula closed to those which may be found in the literature, 
we introduce two new positive variables $u$ and $v$ defined by:
\debut
\rho=1-u^2-v^2 \quad {\rm and}\quad z=v^2-u^2 \non
\fin
The domain of validity of the metric is $u\geq0$, $v\geq0$ and $u^2+v^2\leq1$. 
Simple computations using eqs.(\ref{sig0},\ref{g22},\ref{g11})
lead to the following line element:
\debut
ds^2&=&-2\frac{(1-u^2-v^2)^{\frac{3}{2}}}
{\sqrt{1-u^2}\sqrt{1-v^2}(uv+\sqrt{1-u^2}\sqrt{1-v^2})^2}\, dudv\label{kahn}\\
&&+(1-u^2-v^2)\left(\frac{1+u\sqrt{1-v^2}+v\sqrt{1-u^2}}{1-u\sqrt{1-v^2}-v\sqrt{1-u^2}}
\, dx^2+\frac{1-u\sqrt{1-v^2}-v\sqrt{1-u^2}}{1+u\sqrt{1-v^2}+v\sqrt{1-u^2}}\, 
dy^2\right)\non
\fin
This is the Khan-Penrose solution. 

Another class of well-known solutions are provided by Kasner's solutions. 
They correspond to the parameters $\{(z_p,u_p)\}=\{(z,u)\}$ 
in the limit $z\rightarrow+\infty$.
Equivalently they correspond to the vertex operator
$W_{u}(w)$ with $w\rightarrow1$. 
With $\displaystyle{t=\rho^{\frac{u^2+3}{4}}}$ 
and after a correct redefinition of the variables in order to absorb null factors 
and non significative constants we deduce that:
\debut
ds^2&=&-dt^2+t^{2p_1}dz^2+t^{2p_2}dx^2+t^{2p_3}dy^2\non
\fin
with $\displaystyle{p_1=\frac{u^2-1}{u^2+3}}$, $\displaystyle{p_2=2\frac{1-u}{u^2+3}}$ and $\displaystyle{p_3=2\frac{1+u}{u^2+3}}$ such that 
$p_1^2 + p_2^2 + p_3^2 =1$ and $p_1 + p_2 + p_3 =1$.

\subsection{Non-diagonal solutions.}
In this subsection, we will focus on two examples of solutions describing 
collisions of two impulsive gravitational waves with non-colinear polarization vectors: 
the Chandrasekhar-Xantopoulos solution \cite{ChanXan} and its dual, 
the Nuktu-Halil solution \cite{Nuktu}, and the more general Ernst family 
of solutions \cite{Ergarcia}. 
To obtain more familiar and more compact results,
we shall once more introduce two new variables $\xi$ and $w$.
They correspond to the prolate spheroidal coordinates.  
Details on the relation between these various coordinates and
related identities may be found in Appendix C.

The Chandrasekhar-Xantopoulos solution is generated 
by the following sets of parameters:
\debut
\{(z_p,u_p)\}=\left\{(1,-1),(+\infty,-1),(-1,-1)\right\}
\quad {\rm and}\quad
\{(z_j,y_j)\}=\left\{\left(1,-\frac{q}{4(1+p)}\right),
\left(-1,\frac{q}{4(1+p)}\right)\right\}\non
\fin
where $p^2+q^2=1$. Here $p,q$ are simply used to parametrize 
$y_1=-y_{-1}$. The corresponding vertex operator is:
\debut
g&=&W_{-1}(\infty)\, W_{-1}(1)\, W_{-1}(0)\, 
\Bigl({1-i\frac{q}{4(p+1)}W_2(\infty)}\Bigr)\, 
\Bigl({1+i\frac{q}{4(p+1)}W_2(0)}\Bigr)\non
\fin
Einstein's equations allow to adjust metrics elements. In particular,
we can add a constant to the imaginary part of 
the Ernst potential or multiply it by a global constant factor. 
Using this freedom to translate the imaginary part by $-\frac{2q}{1+p}$, 
we obtain for the Ernst potential:  
\debut
\frac{\rho}{G_{22}} + i \frac{G_{12}}{G_{22}} &=&
\frac{\vev{Wg^*}_{z,\rho}}{\vev{g^*}_{z,\rho}} \label{chand}\\ 
&=&\frac{\rho\left((1-p\xi)^2+q^2w^2\right)-iq\left(-2\xi(1-w^2)+2p(\xi^2-w^2)+
\frac{1}{1+p}(p^2(w^2-\xi^2)+(1-w^2))\right)}{1-p^2\xi^2-q^2w^2}\non
\fin
The conformal factor is determined by computing the expectation
value $\vev{g}_{z,\rho}$.
With $X=(1-p\xi)^2+q^2w^2$ and $Y=1-p^2\xi^2-q^2w^2$, the line element can be written as:
\debut
ds^2&=&-\frac{X}{2}\left(\frac{d\xi^2}{1-\xi^2}-\frac{dw^2}{1-w^2}\right)
+\frac{Y}{X}\left(dx-\frac{2q}{p(p+1)}dy\right)^2
+\frac{4q(1-w^2)(1-p\xi)}{pX}\left(dx-\frac{2q}{p(p+1)}dy\right)dy\non\\
&&+\frac{(1-w^2)}{p^2X}\Bigl({(1-p\xi)^2+q^2)^2+p^2q^2(1-\xi^2)(1-w^2)}\Bigr)dy^2\non
\fin
This is the Chandrasekhar-Xantopoulos solution \cite{ChanXan} written in
the same form as in \cite{Grif}.

To illustrate the duality formulas, we compute the dual of this metric. 
It is the Nuktu-Halil solution \cite{Nuktu}. 
The dual vertex operator is :
\debut
g^* = W_1(\infty)\, W_1(0)\, \Bigl({1-i\frac{q}{4(p+1)}W_{-2}(\infty)}\Bigr)\,
\Bigl({1+i\frac{q}{4(p+1)}W_{-2}(0)}\Bigr)\non
\fin
The dual Ernst potential is then given by:
\debut
\frac{\rho}{G^*_{22}} + i \frac{G^*_{12}}{G^*_{22}} = 
\frac{\vev{Wg}_{z,\rho}}{\vev{g}_{z,\rho}} 
=\frac{1+p\xi+iqw}{1-p\xi-iqw}\label{nuk}
\fin
from which the metric can be computed.

These solutions may be seen as members of a larger class of solutions, 
the so-called Ernst family of solutions \cite{Ergarcia}. 
These correspond to the following 
sets of parameters:
\debut
\left\{(1,-1),(+\infty,-n),(-1,-1)\right\}\quad {\rm and}\quad
\left\{\left(1,\frac{q-q'}{4(p+p')}\right),
\left(-1,\frac{q+q'}{4(p+p')}\right)\right\}\non
\fin
with $p^2+q^2=1$ and $p'^2+q'^2=1$. Introducing $p,q$ and $p',q'$ is just
a convenient  way of parametrizing $y_1$ and $y_{-1}$.
The vertex operators are:
\debut
g&=&W_{-n}(\infty)\, W_{-1}(1)\, W_{-1}(0)\, 
\Bigl({1+i\frac{q-q'}{4(p+p')}W_2(\infty)}\Bigr)\, 
\Bigl({1+i\frac{q+q'}{4(p+p')}W_2(0)}\Bigr)\non
\fin
The expression of the metric is a little lengthy.
For the conformal factor we have
 \debut
e^{2\hat\sig}\, (d\rho^2-dz^2)&=& |\vev{g}_{z,\rho}|^2\, (d\rho^2-dz^2) \non\\
&=&\frac{1}{4}\rho^{\frac{n^2}{2}}
\left(\frac{d\xi^2}{1-\xi^2}-\frac{dw^2}{1-w^2}\right)
\left[{(1-\xi^2)((p+p')^2(\frac{1-\xi}{1+\xi})^n
+(p'-p)^2(\frac{1+\xi}{1-\xi})^n)}\right.\non\\
&&+\left.{(1-w)^2((q+q')^2(\frac{1-w}{1+w})^n
+(q'-q)^2(\frac{1+w}{1-w})^n)+2(q^2-q'^2)(\xi^2-w^2)}\right]\non
\fin
For the Ernst potential we have
\debut
\frac{\rho}{G_{22}} + i \frac{G_{12}}{G_{22}} =\frac{\vev{Wg^*}_{z,\rho}}{\vev{g^*}_{z,\rho}}=\rho^n\frac{A}{B}\non
\fin
with
\debut
A&=&(1-\xi^2)^\half\left[(p+p')\left({\frac{1-\xi}{1+\xi}}\right)^{\frac{n+1}{2}}
+(p'-p)\left({\frac{1+\xi}{1-\xi}}\right)^{\frac{n+1}{2}}\right]\non\\
&&+i(1-w^2)^\half\left[(q+q')\left({\frac{1-w}{1+w}}\right)^{\frac{n+1}{2}}
+(q'-q)\left({\frac{1+w}{1-w}}\right)^{\frac{n+1}{2}}\right]\non
\fin
and
\debut
B&=&(1-\xi^2)^\half\left[(p+p')\left({\frac{1-\xi}{1+\xi}}\right)^{\frac{n-1}{2}}
+(p'-p)\left({\frac{1+\xi}{1-\xi}}\right)^{\frac{n-1}{2}}\right]\non\\
&&+i(1-w^2)^\half\left[(q+q')\left({\frac{1-w}{1+w}}\right)^{\frac{n-1}{2}}
+(q'-q)\left({\frac{1+w}{1-w}}\right)^{\frac{n-1}{2}}\right]\non
\fin
These are the solutions found in \cite{Ergarcia}.
The dual metrics can be obtained by changing $n$ into $1-n$ 
and exchanging $(p,q)$ with $(p',q')$. This may be checked 
by comparing the dual vertex operators.

\subsection{Analytic continuation.}
Stationary axis symmetric solutions of the vacuum Einstein equations
may formally be obtained by analytic continuation:
\debut
\rho\to i\rho\ , \ x\to i\varphi\ ,\
z\to z\ ,\ y\to i\tau \label{anal}
\fin
The (dual) metric then reads:
\debut
ds_*^2 = - \De^{-1}(d\tau + \om d\varphi)^2 + \De e^{2\hat \sig}(dz^2 + d\rho^2)
+ \rho^2 \De d\varphi^2 
\label{axis}
\fin
with $\om=N^*$. However the reality conditions for axis symmetric solutions
are more involved.  

By analytic continuation, the Khan-Penrose solution 
is mapped into the Schwarzchild solution.
Let's see what happen in the case of the Chandrasekhar-Xantopoulos solution. 
Using the standard parametrisation  recalled in Appendix C, cf eg.\cite{Grif},
one obtains the line element of the Kerr solution: 
\debut
2M^2ds^2&=&-\left(1-\frac{2Mr}{R^2}\right)d\tau^2+
\frac{4aMr}{R^2}\sin^2{\theta}{d\tau}d\phi\non\\
&&+\left(r^2+a^2-\frac{2a^2Mr}{R^2}\right)\sin^2\theta{d\phi}^2
+R^2\left(\frac{1}{D}dr^2+d\theta^2\right)\non
\fin
with $R^2=r^2+a^2 \cos^2\th$ and $D=r^2-2Mr+a^2$.
Here the domain of validity of the metric is such that $-(M^2-a^2)<D\leq0$,
which corresponds to the region inside the ergo-sphere.
However, using the analytic continuation described above, eq.(\ref{anal}),
the variables $\xi$ and $w$ become
\debut
\xi&=&\frac{1}{2}\left(\sqrt{(1-z)^2+\rho^2}+\sqrt{(1+z)^2+\rho^2}\right)\non\\
w&=&\frac{1}{2}\left(\sqrt{(1-z)^2+\rho^2}-\sqrt{(1+z)^2+\rho^2}\right)\non
\fin
The domain is now such that $D\geq0$. 
The solution thus describes the asymptotically flat exterior Kerr solution,
which is stationary axis symmetric.

Other physically realistic axis symmetric solutions would probably require
infinite sets of parameters $(z_p,u_p)$ and $(z_j,y_j)$, ie. 
infinite products of vertex operators. In such a case the metric coefficients
could be expressed in terms of Fredholm determinants. However the analysis
of such cases is beyond the scope of this paper.

\subsection{Belinskii-Zakharov approach : one soliton case.}
To make contact with the Belinskii-Zakharov approach, 
we describe in more detail the vertex operator construction
of the one soliton solution found in \cite{Zakha}.
The Belinskii-Zakharov approach \cite{Zakha} starts
from Kasner's metric as a seed solution. Since these
Kasner's solutions are obtained from the vertex operator with
parameter $\{(z_p,u_p)\}=\left\{(+\infty,u)\right\}$,
we insert vertex operators whose sets of parameters are now
\debut
\{(z_p,u_p)\}=\left\{(+\infty,u),(-1,-1)\right\}
\quad {\rm and}\quad
\{(z_j,y_j)\}=\left\{\left(-1,y\right)\right\}\non
\fin
They correspond to
\debut
g= W_u(1)W_{-1}(0)\ (1+iyW_2(0)) \non
\fin
Given these values of the parameters $z_j$ one has to compute 
the values of $\mu_j$ using eq.(\ref{defY}). To compare with
ref.\cite{Zakha} let us introduce the same notation as this reference:
\debut
e^{\frac{r}{2}} \equiv \frac{1-\mu_{-1}}{1+\mu_{-1}} 
 =  \frac{(z+1)}{\rho}-\sqrt{\frac{(z+1)^2}{\rho^2}-1}\non
\fin
Using this parametrization and appropriate normalization,
we find using formulas of Section 1 the metric:
\debut
ds^2&=&\frac{\rho^{2q^2}\cosh\left(qr+C\right)}{\sqrt{(z+1)^2+\rho^2}}
	\left(-d\rho^2+dz^2\right)
	+\frac{\cosh\left(\left(\frac{1}{2}+q\right)r+C\right)}{\cosh\left(qr+C\right)}
	\rho^{1+2q}dx^2 \non\\
	&&-\frac{2\sinh\left(\frac{r}{2}\right)}{\cosh\left(qr+C\right)}
	\rho dxdy
	+\frac{\cosh\left(\left(\frac{1}{2}-q\right)r-C\right)}{\cosh\left(qr+C\right)}
	\rho^{1-2q}dy^2\non	
\fin
with $e^C=4y$ (choosing $y$ positive) and $u=2q$.
This coincides with the one soliton solution of Belinskii and Zakharov.

So the formulas we have found can be used in the same way as those found 
following the Belinskii-Zakharov method. 
One may first consider a seed solution and then insert 
more vertex operators to generate multiple solitons solutions. 
As pointed out in Section 2, this leads to solutions
similar to those found by P. Letelier \cite{letel}
but expressed and parametrized in different ways and
with no quadrature left.

The parameters $\mu_j$ defined in eq.(\ref{defY}) correspond 
to the moving poles and the parameter $y$ to the integration
constants of Belinskii and Zakharov. The relation between
the parameters of the vertex operators and the moving poles
can be made more explicit. Namely, let $z_j$ the positions
of the vertex operators and $\mu_j$ defined as in eq.(\ref{defY}) by
$\mu_j^2=\frac{z_j-z+\rho}{z_j-z-\rho}$, then the functions
\debut
\la_j\equiv \rho\left(\frac{1-\mu_j}{1+\mu_j}\right)
= (z-z_j)-\sqrt{(z-z_j)^2-\rho^2}\label{moving}
\fin
are the moving poles of the Belinskii-Zakharov method.
However the main difference between the vertex operator and
the Belinskii-Zakharov approaches is the fact that
no quadrature is needed to obtain the metric in the
vertex operator approach.

\vskip 1.5 truecm

{\bf Acknowledgement:}
It is pleasure to thank Vincent Pasquier for useful discussions.

\vskip 1.5 truecm

\section{Appendix A: A survey of the general method.}
To make the paper more self-contained, we recall 
here some of the results obtained in \cite{BeJu}. 
They could be helpful to understand the general method
based on vertex operators. There are a few steps which we now explain.
We will use notations and formulas introduced in Section 3 and Section 5.
In particular we need to use the Kac-Moody and Virasoro algebras
defined in Section 5.1, eq.(\ref{comh}) and below.

Let us first introduce the Lax connection:
\debut
A_{\pm}=\pm d_{\pm}(L_0-L_{\pm 1})+Q_{\pm}+
P_{\pm}\otimes t^{\pm 1}\mp(\partial_{\pm}\hat\sig)\frac{k}{2}\non
\fin
It is such that its zero curvature condition 
$\left[\partial_++A_+,\partial_-+A_-\right]=0$
is equivalent to $d_{\pm}=\rho^{-1}\partial_{\pm}\rho$
with $\partial_+\partial_-\rho=0$
and the reduced Einstein equations.
As usual the zero curvature condition is the compatibility condition
for an auxiliary  linear system:
\debut
(\partial_{\pm}+A_{\pm})\Psi=0   \label{auxi}
\fin
The solution $\Psi$ of that system is called the wave function.

The simplest solution to Einstein's equations
corresponds to $Q_{\pm}=P_{\pm}=\hat{\sigma}=0$.
We call it the vacuum solution. It is easy to realise that it
is associated to Minkowski's flat solution (using the dual 
metric and analytic continuation). For that solution the
Lax connection is simply $A_\pm= \pm d_{\pm}(L_0-L_{\pm 1})$.
Its wave function is:
\debut
\Psi_0(u,v)&=&\left(\frac{b(v)+c_1}{\rho}\right)^{L_0-L_1}
	\left(\frac{b(v)+c_1}{c_2}\right)^{L_0-L_{-1}} \non\\
	&=&\left(\frac{\rho}{a(u)+c_3(c_1,c_2)}\right)^{L_0-L_{-1}}
	\left(\frac{c_4(c_1,c_2)}{a(u)+c_3(c_1,c_2)}\right)^{L_0-L_1} \non
\fin
with $\rho=a(u)+b(v)$ and $z=a(u)-b(v)$ as in eq.(\ref{rhozuv}) and
where $c_1,c_2$ are constants depending on the initial conditions. 
In this paper, we take $c_1=c_3=\frac{1}{2}$ and 
$c_2=c_4=1$ (choosing other values amounts rescaling and
translating $z$ and $\rho$).

The algebraic vertex operator method is based on manipulating
the wave function to generate new solutions from old ones.
These manipulations are dressing transformations \cite{sem}. 
They were applied to 2D reduced Einstein equations in ref.\cite{BeJu}. 
Dressing symmetries are associated to the factorization problem
in the Kac-Moody group recalled in Section 5.1, eq.(\ref{facto}).
The point is that given the vacuum wave function $\Psi_0$ and 
any element $g=g_-^{-1}g_+$ of the affine $SL(2,R)$
Kac-Moody group with the factorisation
$g_{\pm}\in \exp{\left(\CB_{\pm} \oplus \C k\right)}$,
then the wave function
\debut
\Psi = \left(\Psi_0 g \Psi_0^{-1}\right)_- \Psi_0 \, g^{-1}_-
	= \left(\Psi_0 g \Psi_0^{-1}\right)_+ \Psi_0\,  g^{-1}_+ \non
\fin
is a solution of a compatible auxiliary linear system (\ref{auxi}).

Given a wave function $\Psi$, 
the original fields $Q_\pm, P_\pm$ and $\phi,\hat \sig$,
solutions of the Einstein equations, are then reconstructed by evaluating
matrix elements of the wave function. This is done with the
help of vertex operators. Choosing as in ref.\cite{BeJu}
highest weight representations of the Kac-Moody group 
in which the group elements $g$ can be written 
as products of vertex operators one recovers formulas 
(eq.(\ref{tau0},\ref{tau+},\ref{tau-})) for the Lax connection,
and more interestingly formulas for the elements of the metric (eq.(\ref{magic3}). 
(More details may be found in ref.\cite{BeJu}.)
We mark that in formula (\ref{magic3}) 
all the coordinates' dependence is contained in the vacuum 
wave function $\Psi_0$. 

Note that in this approach the Lax connection does
contain any space-time dependance spectral parameter. 
There is no moving spectral parameter. In particular the poles, 
which are the $w$-argument of the vertex operators, are fixed 
in contrast with other methods based on the Belinskii-Zakharov approach. 
The space-time dependence comes back when we conjugate 
the vertex operators with $\Psi_0$, cf eq.(\ref{wmuAB}). 
This is manifest in
the definition (\ref{moving}) of the moving pole
which may be rewritten as:
\debut
\la = \rho \Psi_0\ \({\frac{1-w}{1+w}}\)\ \Psi_0^{-1} \non
\fin 
So the Virasoro algebra appears as a way to encode the coordinates' 
dependence of the moving poles.

\section{Appendix B: Vertex operator expectation values.}
Here we have gathered a few formulas for the vertex 
operator expectation values.
These are computed using Wick's theorem:
\debut
\vev{\prod_p W_{u_p}(\mu_p)} =
\prod_{p<q}\({\frac{\mu_p-\mu_q}{\mu_p+\mu_q}}\)^{u_p\cdot u_q/2}\label{wick}
\fin
To evaluate the tau-function (\ref{tau0}) or its dual one
needs to compute expectation values such as
$$
\vev{\prod_{p=1}^mW_{u_p}(\mu_p)\cdot 
\prod_{j=1}^n\({ 1 + i Y_j W_2(\mu_j) }\) }
$$
This is done using Wick's theorem (\ref{wick}). For example 
\debut
\vev{\prod_{j=1}^n(1+iY_j W_2(\mu_j))}= \sum_{p=0}^n i^p
\sum_{k_1<\cdots<k_p} Y_{k_1}\cdots Y_{k_p}
\prod_{k_i<k_j} \({\frac{\mu_{k_i}-\mu_{k_j}}{\mu_{k_i}+\mu_{k_j}}}\)^2
\non
\fin
The determinant formula for the tau-functions then follows
from the Cauchy determinant formula
\debut
\det\({ \frac{2\mu_i}{\mu_i+\mu_j} }\) = 
\prod_{i<j}\({\frac{\mu_{i}-\mu_{j}}{\mu_{i}+\mu_{j}}}\)^2 \non
\fin
To evaluate the tau-functions (\ref{tau+},\ref{tau-}) one needs
to know
\debut
\vev{\prod_j W_{u_j}(\mu_j)\cdot p_{-1}}
&=& \prod_{i<j} \({\frac{\mu_{i}-\mu_{j}}{\mu_{i}+\mu_{j}}}\)^{u_i\cdot u_j/2}
\cdot \sum_j u_j~ \mu_j^{-1} \non\\
\vev{p_1\cdot\prod_j W_{u_j}(\mu_j)} 
&=& -\prod_{i<j} \({\frac{\mu_{i}-\mu_{j}}{\mu_{i}+\mu_{j}}}\)^{u_i\cdot u_j/2} 
\cdot \sum_j u_j~ \mu_j \non
\fin
from which equations (\ref{vgp}) follow.

\section {Appendix C: Some details about the examples}
Here we have gathered a few formulas and identities that can be helpful 
in the computations used in the examples.
First, let introduce the two variables $\xi$ and $w$ by
\debut
\xi&=&\frac{1}{2}\left(\sqrt{(1-z+\rho)(1-z-\rho)}
+\sqrt{(1+z+\rho)(1+z-\rho)}\right)\non\\
w&=&\frac{1}{2}\left(\sqrt{(1-z+\rho)(1-z-\rho)}
-\sqrt{(1+z+\rho)(1+z-\rho)}\right)\non
\fin
Reciprocally
\debut
\rho&=&\sqrt{(1-\xi^2)(1-w^2)}\non\\
z&=&-w\xi\non
\fin
The domain of definition of $\rho$ and $z$ implies that $|w|\leq\xi\leq1$.

We use the notation $\mu_z$ and $X_z$ for the variables $\mu$ and $X$
associated to $z$ as defined in eq.(\ref{defY}). 
The singular values of $z$ we have chosen, eg. $z=\pm 1$ or $z=\infty$,
lead to simple expressions for the factor using $\mu_z$. For example:
\debut
\lim_{z\rightarrow-1} X_z\left(\frac{\mu_z-\mu_{-1}}{\mu_z+\mu_{-1}}\right)^{-1}
=\lim_{z\rightarrow1} X_z\left(\frac{\mu_1-\mu_z}{\mu_1+\mu_z}\right)^{-1}=4\non
\fin
and
\debut
\left(\frac{\mu_z-1}{\mu_z+1}\right)\sim\frac{\rho}{2z}\
\quad {\rm for}\quad z\rightarrow+\infty\non
\fin
The variables $\mu_z$ for $z=\pm 1$ are related to the variables $\xi$ and $w$ by
\debut
\left(\frac{\mu_1-1}{\mu_1+1}\right)\left(\frac{1-\mu_{-1}}{1+\mu_{-1}}\right)
=\frac{1-\xi}{1+\xi}\non\\
\left(\frac{\mu_1-1}{\mu_1+1}\right)\left(\frac{1+\mu_{-1}}{1-\mu_{-1}}\right)
=\frac{1-w}{1+w}\non\\
\left(\frac{\mu_1-\mu_{-1}}{\mu_1+\mu_{-1}}\right)
=\left(\frac{1-\xi^2}{1-w^2}\right)^{\half}\non
\fin
The variables $X_z$ for $z=\pm 1$ are related to the variables $\xi$ and $w$ by
\debut
\frac{\left(X_1X_{-1}\right)^\frac{1}{2}}{\rho}\left(d\rho^2-dz^2\right)
&=&{\rm const.}\, \left(\frac{d\xi^2}{1-\xi^2}-\frac{dw^2}{1-w^2}\right)\non
\fin
Finally we give the parametrisation used to deduce the Kerr solution 
from the Chandrasekahar-Xanthopoulos one. 
The relation between the parameters $(p,q)$ and the mass 
and the angular momentum is
\debut
p=-\frac{\sqrt{M^2-a^2}}{M}\quad ;\quad q=\frac{a}{M}\non
\fin
The relations between the coordinates are
\debut
\xi=\frac{r-M}{\sqrt{M^2-a^2}}\quad&;&\quad w=\cos\theta\non\\
\tau=-\sqrt{2}M\left(x-\frac{2q}{p(1+p)}y\right)\quad &;&\quad\phi
=\frac{\sqrt{2}M}{\sqrt{M^2-a^2}}y\non
\fin
Recall that $R^2=r^2+a^2\cos^2\th$ and $D=r^2-2Mr+a^2$.

\newpage

\end{document}